\documentclass[aps,pre,twocolumn,showpacs,amsmath,amssymb]{revtex4}
\usepackage{graphicx} 
\usepackage{dcolumn} 
\usepackage{bm} 

\begin{document}

\title{Geometric control of failure behavior in perforated sheets}
\author{Michelle M. Driscoll}
\affiliation{ The James Franck Institute and Department of Physics, The University of Chicago }

\date{\today}

\begin{abstract}
Adding perforations to a continuum sheet allows new modes of deformation, and thus modifies its elastic behavior.  The failure behavior of such a perforated sheet is explored, using a model experimental system: a material containing a one-dimensional array of rectangular holes.  In this model system, a transition in failure mode occurs as the spacing and aspect ratio of the holes are varied: rapid failure via a running crack is completely replaced by quasi-static failure which proceeds via the breaking of struts at random positions in the array of holes.  I demonstrate that this transition can be connected to the loss of stress enhancement which occurs as the material geometry is modified. \end{abstract}

\pacs{46.50.+a, 62.20.mm,62.20.mt}

\maketitle

\section{Introduction}
\label{intro}

Material failure occurs in many ways: from plastic deformation \cite{tvergaard1990material,garrison1987ductile}, to slowly creeping fatigue damage \cite{fatigue_of_structures, forsyth1963fatigue}, to a sudden and catastrophic (brittle) failure of an entire structure \cite{griffith1921phenomena,bouchbinder2010dynamics}.  This variety of failure modes hints that individual material properties must determine a structure's fate. Fracture mechanics has made much progress in predicting when a structure will start to fail, for example by giving analytic solutions for stress and strain fields created by cracks in simple geometries  \cite{freund, irwin1957analysis}.  However, there are still many open questions. For example, fracture mechanics makes no prediction about what path a running crack will follow; this is an input parameter to the theory \cite{sym}.  Furthermore, fracture mechanics is a  continuum theory and it is not necessarily clear how the predictions it makes map to structures with complex geometries.

One such geometry is a perforated structure,  created by introducing holes into a continuum solid.    Introducing these holes creates a new meta-material that can deform in very different ways, ways that are too energetically costly in a continuum solid.  Thus, these meta-materials will have different material properties, which will depend the perforation spacing and on the geometry of the holes \cite{Day19921031,jasiuk1994elastic,cellular_solids,hu2000effective}.  

\begin{figure}
\centering 
\begin{center} 
\includegraphics[width = \columnwidth]{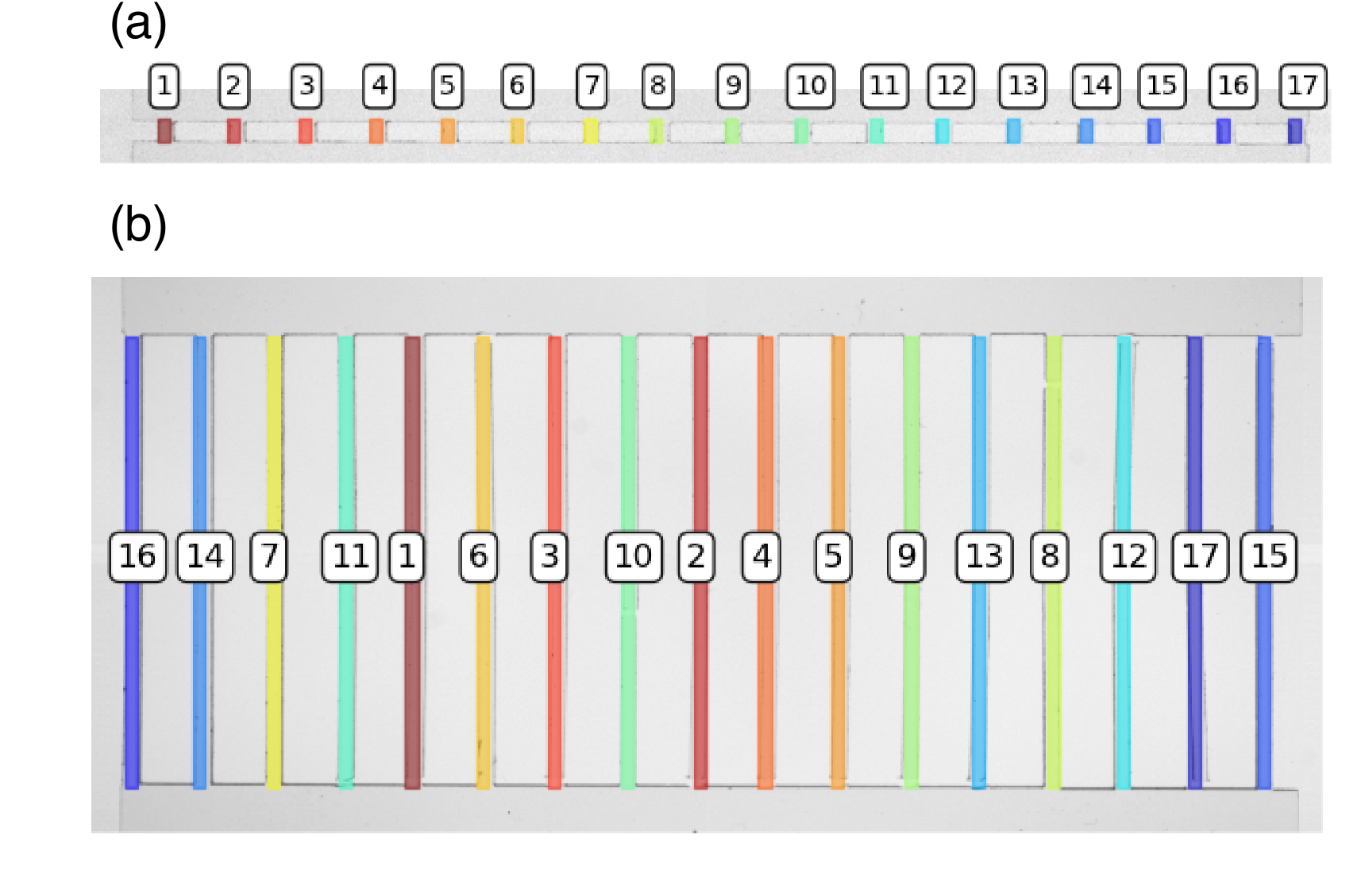} 
\caption{(Color online) When the geometry of the 1D meta-material is changed, the failure mode transitions from a running crack to random breaking; the strut breaking order as well as the total failure time, $\tau$ (time interval between first and last break) changes markedly.  Numbering and color both indicate strut breaking order; colormap runs from red to blue.  (a) When the struts composing the material are small and spaced relatively closely, failure proceeds via a propagating crack: defined as ordered breaking of struts, which occurs at speeds comparable to material sound speeds.  Here, $\tau = 149.3 \mu$s. (b) A transition in failure mode occurs when the struts become long and narrow, or spaced far apart from each other.  In this regime, failure occurs via the breaking of struts in a non-ordered fashion, and the breaking rate is set by the pulling rate, i.e., failure is quasi-static. Here, $\tau = 1.58$ s.} 
\label{phenom}
\end{center}
\end{figure}

In this work, I  study  the failure of perforated meta-materials.  A material with a one-dimensional array of holes is used as a model system to study how geometry modifies failure behavior.  I find that the failure behavior of this one-dimensional system can be characterized in terms of hole spacing and geometry, and a distinct transition in failure mode occurs as the material geometry is modified.  This transition is characterized by a change in failure dynamics as the perforation geometry is altered: failure via a running crack state is replaced by failure via quasi-static, random breaking.  Figure \ref{phenom} illustrates this transition; in the propagating crack regime, failure occurs very rapidly and in an ordered manner, while in the random-breaking regime, failure occurs in a non-ordered manner, and at a timescale set by the displacement rate of the material, i.e.\ quasi-statically.  Furthermore, I show that this transition can be connected with a loss of stress enhancement in the material as geometry is altered.  

Note that in these studies of perforated materials the individual bonds all have nearly identical strengths.  There is no disorder in the system.  This is distinct from other models that have been studied such as the fiber bundle model \cite{herrmann2009some, daniels1945statistical, pradhan2010failure} where the amount of disorder is varied and produces changes in the way in which the material fails.  The present experiments focus specifically on the role of geometry (while keeping the amount of disorder fixed) on the failure mode of the material.

Looking at material failure through the lens of geometry provides a new insight into how a material fails, and provides a direct relationship between a meta-material's microstructure and its overall failure behavior.  Furthermore, these results allow for the possibility of a tunable failure mode --- by simply changing geometry, the same base material can be used to construct a suite of meta-materials with very different failure behaviors.

In the next section, I give a brief overview of the previous work that has been done on fracture in perforated sheets.  Section \ref{methods} outlines the details of the experimental work, as well as the finite element calculations that were done.  Section \ref{1D transition} discusses in detail the transition in dynamics that occurs, and how it is controlled by material geometry.  In the running crack regime, geometry acts to additionally control the velocity of the crack; this is presented in Section \ref{velocity}.

\section{Background on fracture of perforated sheets}

The fracture of perforated solids has been the subject of extensive study, often in the context of measuring material properties such as failure stresses and strains \cite{chen2001effect,huang1991fracture,fleck2007damage,ryvkin2011crack}. For example, perforated geometries have been used as a model system for ductile failure, which is thought to occur via the growth of small voids which then coalesce into a larger defect.  Several experimental and numerical studies have been conducted in perforated metal sheets \cite{dubensky1987void, geltmacher1996modeling,magnusen1990simulation}; the focus of this work has been on measuring or  predicting failure strains and stresses as a function of void fraction or void arrangement in order to gain insight about ductile failure in a continuum material.   Another recent study examined the  failure of an array of perforations in a thin plastic sheet \cite{meunier2013plane}, finding a transition in failure from localized inter-hole failure to large-scale plastic deformation.  This work only examined quasi-static plastic failure, and those results do not generalize to brittle solids.

A propagating crack in a continuum sheet moves at a velocity comparable to material sound speeds; an advancing crack should accelerates as it grows, until it travels at the material's Rayleigh velocity, $c_R$ (the speed of surface waves)\cite{stroh1957theory}.  However, in practice, it is often found that  cracks only accelerate up to a velocity $\sim 0.5 c_R$ before exhibiting  branching  \cite{sharon1996microbranching, ravi1984experimental} or other dynamic instabilities \cite{PhysRevLett.98.124301,deegan2001oscillating,fineberg1999instability}.  However, if these instabilities are suppressed, cracks are observed to move at speeds very near to the Rayleigh velocity \cite{PhysRevLett.98.124301} in continuum sheets.  

In comparison to the large body of work done in continuum solids, very few studies have measured crack dynamics in a perforated material.  The velocity of  a running crack in a perforated material   has been measured in least one study \cite{washabaugh},  but this work focused on examining the behavior of fracture at weak interfaces.  A line of perforations was used as a model for a weak interface, in order to compare the results in the perforated geometry with results obtained in materials with a weak plane.  A dependance of crack velocity on hole area fraction was found, but as the perforated geometry was not the main focus of this work, only a very small set of hole geometries was tested, and only one material was used.  Here, I show that in a perforated material, geometry acts a control parameter for the crack velocity, independent of material properties.

\section{Methods}
\label{methods}

Two approaches were used to study failure in a perforated material.  The bulk of the work was experimental; samples were put under tension until failure, and the resulting failure behavior was analyzed in terms of the material geometry.  To complement the experimental work, finite-element calculations were done to measure elastic properties, as well as to gain insight into how stress fields are modified by geometry.

\subsection{Experiments}
\label{expt}

The experimental samples were fabricated from thin (0.75 mm - 1.5 mm thick) sheets of plastic.  Thin sheets were used to approximate a two-dimensional solid geometry.  The bulk of the work was conducted using 1.5 mm thick cast acrylic sheets purchased from McMaster-Carr, which had  Young's modulus, $Y$ = 3.7 GPa, Poisson's ratio, $\nu$ = 0.35, and density, $\rho$ = 1157 kg/m$^3$.  A small selection of control tests were performed using two additional plastics, 0.75 mm thick  Delrin 150 sheets ($Y$ = 3.1 GPa, $\rho$ = 1394 kg/m$^3$, $\nu$=0.4) and 0.75 mm thick impact-modified acrylic sheets ($Y$ = 1.76 GPa, $\rho$ = 1115 kg/m$^3$, $\nu$=0.4).

\begin{figure}
\centering 
\begin{center} 
\includegraphics[width=\columnwidth]{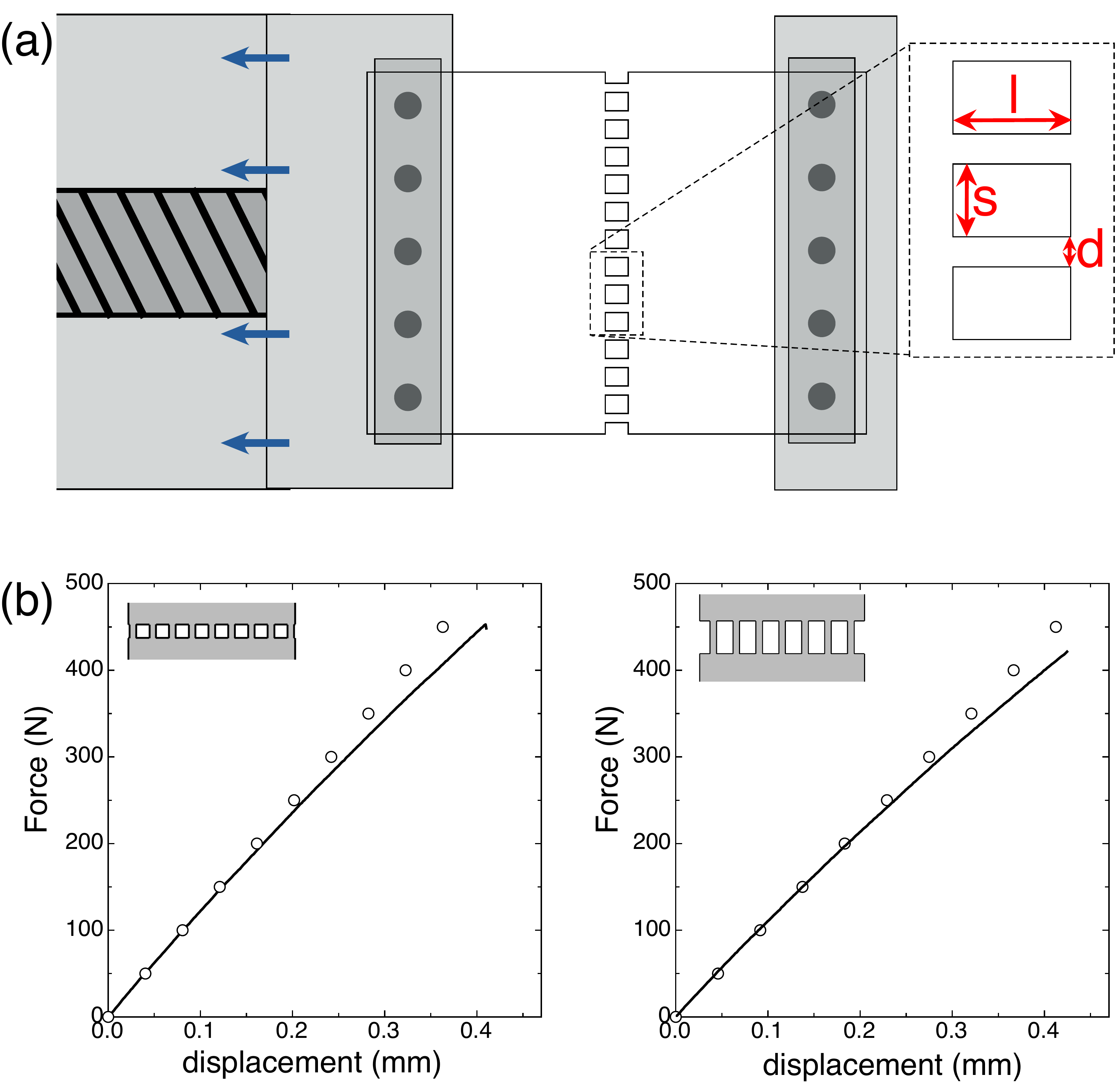} 
\caption{(Color online) (a) Imaging studies were conducted using a custom-built apparatus.  The samples are held fixed at one boundary, while the other boundary is displaced at a constant rate.  Enlargement indicates the parameters that characterize the hole size ($s,l$) and spacing ($d$). (b) Force-displacement curves for two sample geometries, $d$ = 1mm, $s$ = 2 mm, $l$ = 2 mm (left) and  $d$ = 1mm, $s$ = 2.5 mm, $l$ = 5 mm (right).  The lines show the data obtained from the experimental samples, while the open symbols show the results of finite-element calculations.  The good agreement between measurements and calculations demonstrates that the samples behave in a nearly linear fashion.    } 
\label{setup_F_d}
\end{center}
\end{figure}

The samples were fabricated in a ladder geometry, consisting of a 1D array of struts of width $d$ and length $l$, separated by a spacing of $s$, as illustrated in Fig.\ \ref{setup_F_d}a.  Thus, a family of 1D geometries could be constructed by varying the ratio $s/d$ from 0.4 - 50 and the ratio $l/d$ from 0.2 - 79.5.  Additionally, a set of control experiments was run where the absolute value of $d$ was varied from 0.6 mm - 4.7 mm.   The holes in the laser cut samples are not perfectly rectangular due to the finite resolution of the laser cutter ($\sim$ 50 $\mu$m - 100 $\mu$m); the corners are slightly rounded with a radius of curvature $\sim$ 140 $\mu$m.  However, it is not believed that this small amount of  rounding influenced the results, as experimental stress and strain measurements produced good agreement with finite-element calculations which did not have these slightly rounded corners.

As shown in Fig.\ \ref{setup_F_d}a, the samples have tabs on two sides so they could be clamped to a materials tester for the failure studies.  The size of these tabs does not impact the results; most of the strain occurs in the perforated part of the samples.  The effect of tab length was explicitly tested by varying the tab length by a factor of 2; this produced no change in the results.

All samples were broken under uniaxial tension, often termed Mode I failure; the displacement direction is defined as the $\hat{y}$ direction.  A custom apparatus was constructed for displacement controlled failure experiments  as shown in Fig.\ \ref{setup_F_d}a.  One end of the sample is  held on a fixed stage, machined from cast aluminum (MIC 6), while the other end  is held on translation stage made from a cast aluminum block firmly attached to a machine vise.   A geared down DC gearmotor was used to drive this translation stage at a constant displacement rate of 83.0 $\pm 0.5 \mu$m/s.  

Extreme care was taken to ensure that the fixed stage was, to the nearest mil, machined to be the exact height as the translation stage.  This was done to ensure the samples would experience an  even loading, so the failure behavior could be assumed to occur under pure uniaxial tension.  As illustrated in Fig.\ \ref{setup_F_d}a, the samples are attached to both the translation stage and the fixed stage via cast aluminum bars.  Screws were used to apply pressure to the bars, clamping the samples down. Again, care was taken to machine both the surfaces of the stages as well as the bars to be as flat as possible so as not to induce uneven loading conditions.  Failure was observed to initiate on both sides (left and right) of the samples, verifying that they experienced a uniform loading condition.  

A high speed camera (Vision Research, Phantom v1610 and Phantom V9) was used to record the fracture behavior at speeds up to 400,000 fps.  The samples were backlit with a 18W LED light panel (rosco LitePad HO+).  Use of an LED panel assured no significant heat was input to the samples, as significant heating could cause a change in material properties \cite{agrawal2010investigation}.  For some experiments, crossed polarizers (Edmund Optics) were used to  image the stress field of the sample.  This aided in determining the precise location where material failure occurred, as the samples halves only separated by a small amount during the failure process. 

An additional series of tests was performed in a commercial materials tester (Instron 5869) in displacement-controlled mode: a constant displacement is applied to the samples, while the force is measured.  A 50 kN load cell was used for these tests, as the force required to break the samples ranged from 200 N - 1200 N.  Due to the size constraints of the materials tester, the samples fabricated for materials testing were confined to be 25.4 mm or less. 

Figure  \ref{setup_F_d}b  shows the measured force-displacement curves for two sample geometries demonstrating that the curves appear quite linear until failure.  Overlaid on the measured curves are the results of finite element calculations (COMSOL, v4.4) of the force-displacement response for the same geometries.  At small strains, the calculations agree very well with the measurements, emphasizing that these materials remain reasonably close to a linear material, even when near failure.  Slight deviations from the linear behavior become apparent only after the strain has reached half of its value at failure.

\subsection{Finite element calculations}
\label{comsol}

In additional to the experiments, finite element calculations were performed using a commercial package (COMSOL, v4.4).  Simple geometries were constructed directly with the COMSOL interface, to facilitate parameter sweep testing (a suite of studies done by varying one geometric parameter).  The model numerical samples were always made to be identical in size and elastic properties to the experimental samples.  To construct more complex geometries, the CAD files used as laser cutter patterns could be  directly imported into COMSOL.  All calculations were done using the plane stress 2D approximation, as appropriate for a thin plate. 

Two types of calculations were done in the 1D geometry.  To study stress enhancement as a function of geometry,  the constant displacement conditions used in the failure experiments were reproduced: all sample boundaries were free except the upper and lower boundary.  The lower boundary was held fixed and the upper boundary was displaced by a fixed amount corresponding to a strain of 0.005.  Two tests were performed, one in a pristine geometry, and one in which one or more struts were `broken'.  A break was modeled  by manually inserting an additional free boundary (i.e., a cut) in the middle of a strut.  The computed stress field ($\sigma_{yy}$) from both tests was then analyzed to  look for evidence of stress enhancement.  

An additional set of calculations was performed to compare with the Instron test results (Fig.\ \ref{setup_F_d}b).  Here, the lower edge of the sample was held fixed and a constant force was applied to the upper edge of the sample.  This allowed the (maximum) displacement of the sample to be measured and directly compared with the materials-testing results.

\section{One-dimensional fracture: transition in failure mode}
\label{1D transition}

When the holes in the 1D samples are relatively small, and spaced relatively close together, the perforated geometry fails in a manner similar to a solid piece of plastic, via a running crack (as shown in Fig.\ \ref{phenom}a).  Here, the term running crack is used to describe failure that, once initiated,  spreads across the sample at velocities comparable to material sound speeds, with no additional applied strain.   The running crack state is characterized by the struts comprising the sample breaking in rapid succession, and in a directed manner (only nearest neighbor breaks).  

When the struts are long and narrow, or are spaced very far apart, a different mode of failure occurs: random-breaking.  In this regime, the struts do not break in an ordered manner;  many non-nearest-neighbor breaks occur (as shown in Fig.\ \ref{phenom}b).  The breaking rate in this regime is  orders of magnitude slower than in the running crack regime; failure does not spread across the sample after being initiated --- additional failure requires additional applied strain.

\begin{figure}
\centering 
\begin{center} 
\includegraphics[width=\columnwidth]{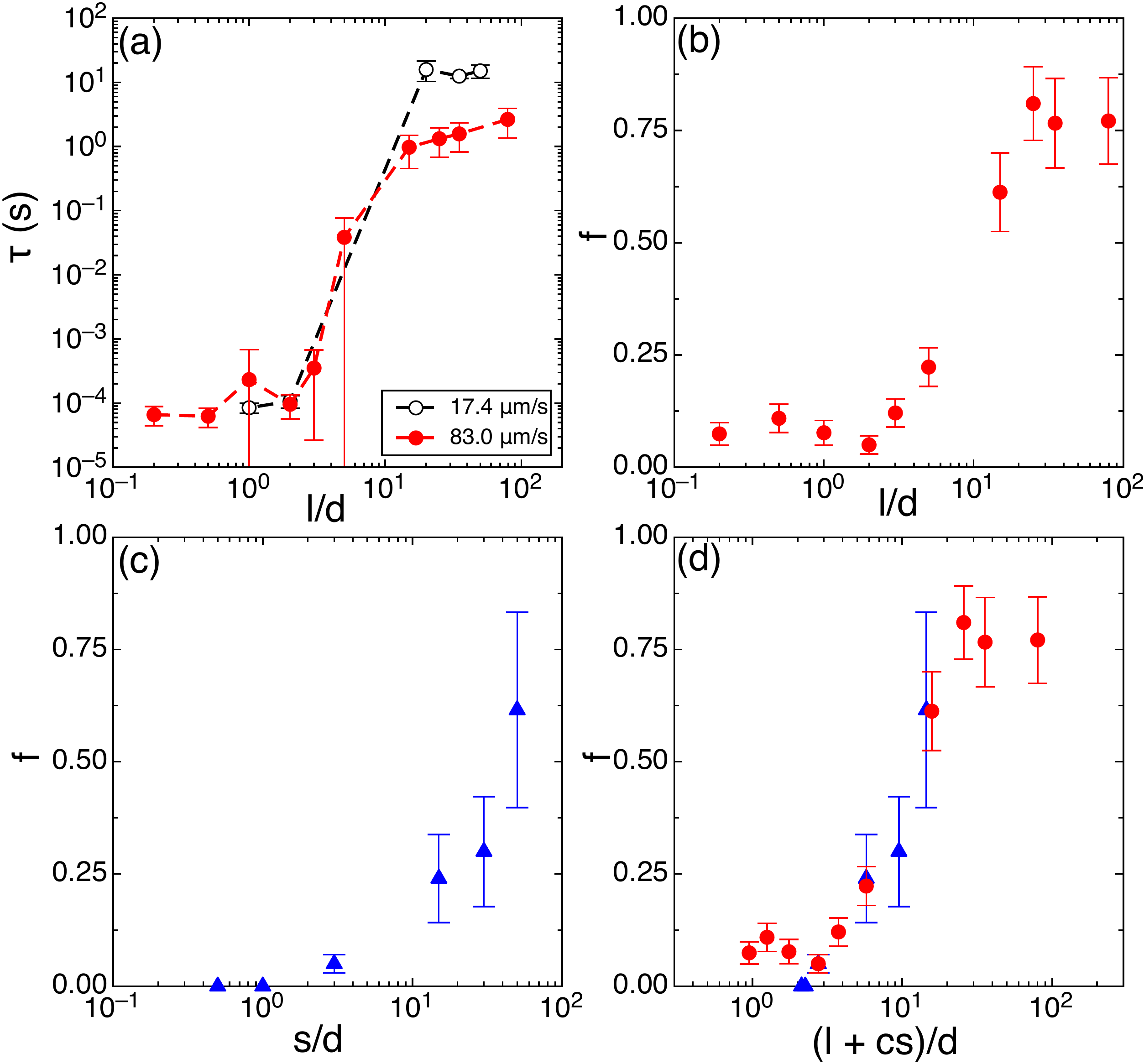} 
\caption{(Color online) Transition in failure mode.   (a) As $l/d$ increases, the failure time $\tau$ increases dramatically.  Data for two displacement rates is shown.  Tests were performed at constant $s/d$ = 3.  (b)  The fraction of nonadjacent breaks, $f$, increases with $l/d$ as well.  (c) $f$ vs.\ $s/d$; tests were performed at constant $l/d$ = 2.  As $s/d$ increases, the fraction of nonadjacent breaks, $f$, increases dramatically. (d) All of the data in (b - circles), (c - triangles) can be collapsed by plotting $f$ as a function of $\frac{l+cs}{d}$, where $c = 0.25 \pm 0.1$.} 
\label{l_d_time_distance}
\end{center}
\end{figure}

To characterize this transition between the two failure modes,  I used two measures: failure time, $\tau$, and the fraction of non-adjacent breaks, $f$.  Failure time is defined as the time for the material to completely break into two separate pieces.  The fraction of non-adjacent breaks, $f$, is defined as the ratio of the number of struts that fail at a position not adjacent to the last strut to fail, normalized by the total number of struts.  Fig.\ \ref{l_d_time_distance}a,b shows how $\tau$ and $f$ dramatically increase as the struts become long and narrow, i.e., as $l/d$ increases.

When $l/d$  is small, failure occurs via a running crack, moving at a velocity comparable to the material sound velocity, thus $\tau$ is very small.  However, as $l/d$ increases, $\tau$ becomes orders of magnitude longer.  The saturation of $\tau$ at high $l/d$ is a reflection of the displacement rate applied to the samples.  To confirm this, identical tests were conducted at a displacement rate that was 5 times slower, $17.4 \pm 0.9 \mu$m/s.  As shown by the open symbols in Fig.\ \ref{l_d_time_distance}a, changing the displacement rate only changes the saturation value at high $l/d$, and does not change the failure time at low $l/d$.

Along with a transition in $\tau$, there is a transition in breaking order as illustrated in Fig.\ \ref{l_d_time_distance}b, which shows the fraction of non-adjacent breaks, $f$,  versus $l/d$.  At small $l/d$, adjacent struts break as a crack runs across the sample.  However, as $l/d$ is increased, this no longer occurs; most breaks are not at adjacent struts, and therefore $f$ increases.  At high $l/d$, $f$ saturates, as there are always some small fraction of adjacent breaks.  Fig.\ \ref{l_d_time_distance}a,b thus illustrates that $l/d$ serves as a control parameter for failure, changing both the failure timescale and breaking pattern of the sample.

However, $l/d$ is not the only control parameter for this transition in the failure dynamics.    Fig.\ \ref{l_d_time_distance}c illustrates that a similar transition in failure behavior is observed as $s/d$ is varied.  As seen in that figure, $f$ increases dramatically as $s/d$ is increased.  As the struts become spaced further and further apart,  failure via a running crack crosses over to random-breaking, similar to what was observed when $l/d$ was increased.

A comparison of $\tau$ as a function of $s/d$ is not shown, as the samples used for this test varied largely in size, thus making comparisons of absolute time difficult.  (Changing $s/d$ experimentally requires making larger and larger samples to ensure a sufficient number of struts in each sample.)  However, $f$ is a normallized quantity, and thus is independent of sample size; it serves as a suitable control parameter to characterize the transition in dynamics as seen in Fig.\ \ref{l_d_time_distance}c.

Figure \ref{l_d_time_distance}a-c illustrates that there are two independent control parameters for the transition from running cracks to random-breaking.  Neither $s/d$ nor $l/d$ alone are sufficient to parameterize the transition; both of these geometric parameters control the failure dynamics separately.  However, Fig.\ \ref{l_d_time_distance}d demonstrates that all of the data can be collapsed onto a single curve by plotting $f$ versus $\frac{l+cs}{d}$, where $c = 0.25 \pm 0.1$.  This rescaling is consistent with the observation that both $l/d$ and $s/d$ control this transition, but in  independent ways.

\begin{figure*}
\centering 
\begin{center} 
\includegraphics[width=2.0\columnwidth]{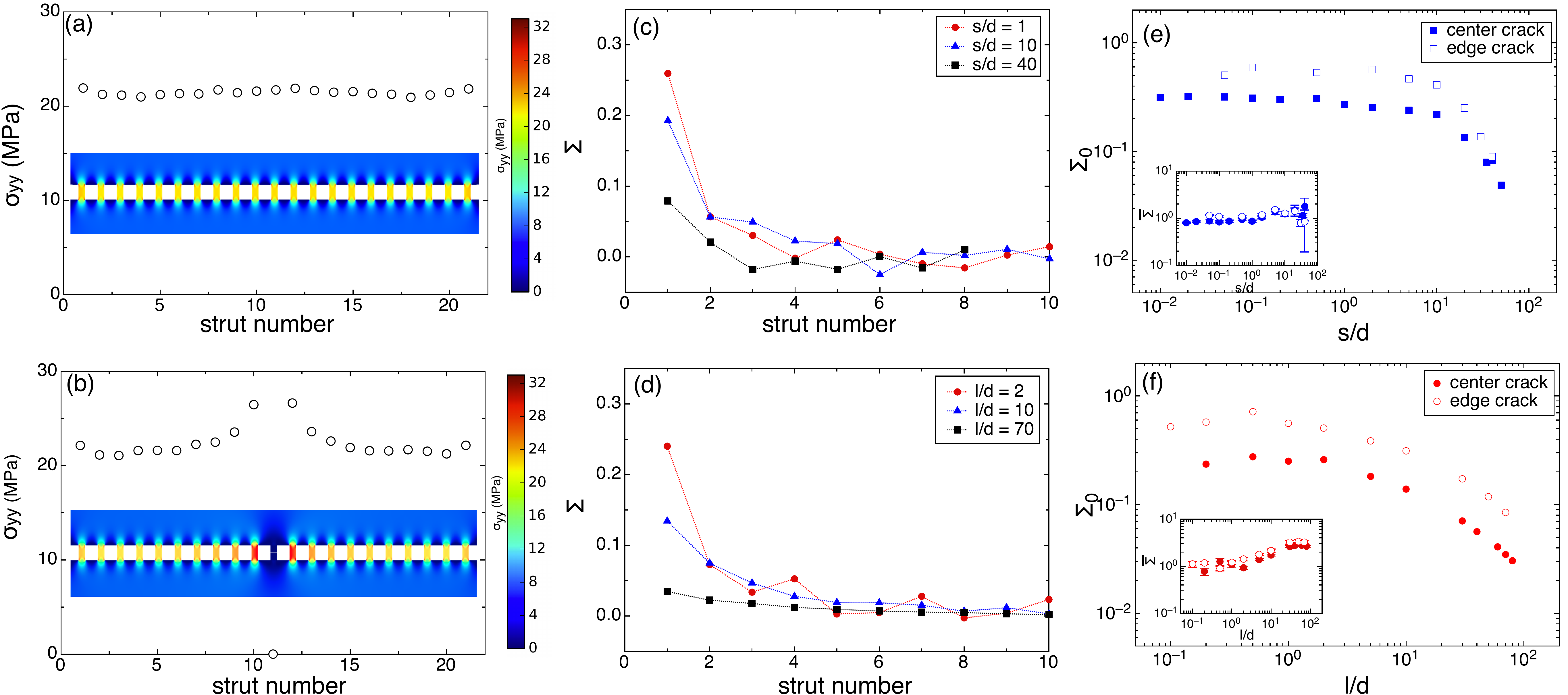} 
\caption{(Color online) Finite element calculations of the stress enhancement due to a broken strut. (a) Stress on each strut ($\sigma_{yy}$) vs.\ strut number; each point represents the total stress on a single strut.  The inset shows a visualization of the stress field in the sample ($\sigma_{yy}$), see color bar for scale. ($\sigma_{yy}$ represents the stress averaged over the strut surface) (b) Stress per strut ($\sigma_{yy}$) vs.\ strut position after the center strut is broken.  There is an enhancement of $\sigma_{yy}$ adjacent to the break location.  The inset shows a visualization of the stress field in the sample ($\sigma_{yy}$), see color bar for scale. (c,d) Normalized stress enhancement, $\Sigma$ vs.\ strut number, for various values of (c) $s/d$  and (d) $l/d$.  The calculations are for a center-broken strut;  points represent the average over the left and right sides of the sample.   $\Sigma$ is a strong function of material geometry.  (e,f) The maximum stress enhancement, $\Sigma_0$  vs.\  (e) $s/d$ and (f) $l/d$.  Data is shown for both crack geometries, a center crack and (filled symbols) an edge crack (open symbols).  In all cases $\Sigma_0$ decreases dramatically as either $s/d$ or $l/d$ is increased.  Insets show the width of the stress enhancement, $\bar{\Sigma}$.  $\bar{\Sigma }$ remains relatively constant as $s/d$ and increases as  $l/d$ is increased.}
\label{new_stress_ex}
\end{center}
\end{figure*}

When a single strut is broken, the stress on the rest of the sample necessarily increases.  More or less of this stress will be borne by the remaining struts, depending on the material geometry.  To gain insight into how the stress enhancement on the struts varies as a function $l/d$ and $s/d$, finite element calculations (COMSOL, v4.4) were performed (for details see Section \ref{comsol}).  A test geometry was input into COMSOL and a fixed displacement was applied.  Then $\sigma_{yy}$ was measured in all of the struts.  Here, $\sigma_{yy}$ is the average value of the yy-component of the stress field in the strut, where the spatial average is computed over the entire surface of the strut.  When no struts are broken, this stress field is relatively uniform across the entire sample, as shown in Fig.\ \ref{new_stress_ex}a.  However, when a break was inserted into the middle strut, an enhancement in stress was found in the remaining struts, as shown in Fig.\ \ref{new_stress_ex}b. This stress-enhancement was observed whether the break was inserted near the center or at one the edge of the sample.

However, this enhancement in stress depends strongly on the material geometry, as shown in Fig.\ \ref{new_stress_ex}c,d.   Here, the normalized  stress enhancement, $\Sigma$, is defined as
\begin{equation}
 \Sigma = \frac{\sigma_{yy}(\mbox{after break}) - \sigma_{yy}(\mbox{before break})}{\sigma_{yy}(\mbox{before break)}},
\end{equation}
and is measured for each strut.  Figure \ \ref{new_stress_ex}c,d presents the measured stress enhancement for two sets of calculations, one in which $s/d$ was varied at fixed $l/d=2$, and one in which $l/d$ was varied at fixed $s/d$ = 3.    As seen in that figure, in all cases, $\Sigma$ is largest near the broken strut, and then decays further away from the break. As either  $s/d$ or $l/d$ is increased, the peak value of $\Sigma$ decreases.

To quantify the overall decrease in stress enhancement shown in Fig.\ \ref{new_stress_ex}c,d, we will use the maximum value,  $\Sigma_0 \equiv \max(\Sigma)$ and the width, $\bar{\Sigma}$, of the stress enhancement.  Fig.\ \ref{new_stress_ex}e,f shows a plot of the maximum stress enhancement, $\Sigma_0$, as $l/d$ and $s/d$ are varied.  Calculations were done in two crack geometries: a center break and an edge break.    

As seen in Fig.\ \ref{new_stress_ex}e,f , in all cases, a transition in $\Sigma_0$ is observed when either $l/d$ or $s/d$ is increased; $\Sigma_0$ is relatively constant, but then begins to drop rapidly.  This transition occurs near $l/d \sim 5$ and $s/d \sim 10$, values which are consistent with the experimentally observed transition in failure mode (see Fig.\ \ref{l_d_time_distance}).   The insets in Fig.\ \ref{new_stress_ex}e,f show  $\bar{\Sigma}$, the width of $\Sigma$ curves, as a function of $s/d$ and $l/d$.  $\bar{\Sigma}$ increases modestly as  $l/d$ is increased, but remains constant as $s/d$ is increased.  The width was determined by fitting a  gaussian to the $\Sigma$ profiles.

\begin{figure}
\centering 
\begin{center} 
\includegraphics[width=\columnwidth]{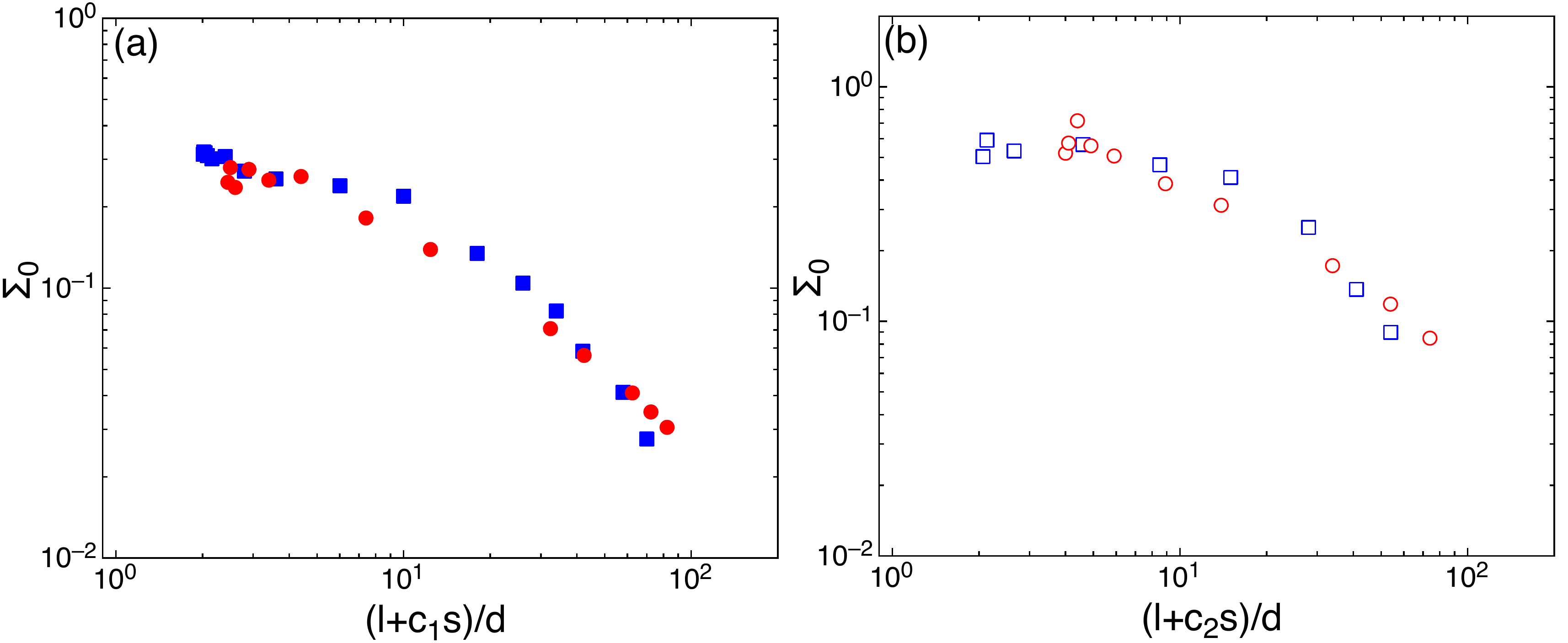} 
\caption{(Color online) The stress enhancement data (Fig.\ \ref{new_stress_ex}e,f) can be collapsed for both crack geometries by plotting $\Sigma_0$ as a function of $\frac{l+cs}{d}$.  (a) $\Sigma_0$ data from the center-crack geometry.  The data was collapsed using  $c_1 = 0.8 \pm 0.1$.   (b) $\Sigma_0$ data from the edge-crack geometry.  The data was collapsed using  $c_2 = 1.3 \pm 0.1$.  } 
\label{comsol_collapse}
\end{center}
\end{figure}

The geometric parameters $l/d$ and $s/d$ separately control the transition between the failure modes,  as demonstrated Fig.\ \ref{l_d_time_distance}.  Likewise, $l/d$ and $s/d$ separately control the stress enhancement present after a strut is broken, as shown in Fig.\ \ref{new_stress_ex}e,f.  This suggests that the data in Fig.\ \ref{new_stress_ex}e,f  can be collapsed in the same manner as the $f$ versus $l/d$ and $f$ versus $s/d$ experimental data.  This collapse is shown in Fig.\ \ref{comsol_collapse}. Here, the stress enhancement data of Fig.\ \ref{new_stress_ex}e,f has been replotted as a function of  $(l+cs)/d$, akin to the collapse used for the experimental measurements.  As seen in that Figure, the data collapses reasonably well for both crack geometries, suggesting that the appearance of the random-breaking failure mode may  be due to a loss of stress enhancement.  Although the parameter $c$  differs slightly in the collapse of the numerical and experimental data, in all cases $c$ is of order one.  Furthermore, it is not clear that $c$ should be the same for all cases, as physically different quantities are being collapsed.  Thus, what is relevant is that the same functional form can be used to collapse all of the data sets, emphasizing that $l/d$ and $s/d$ separately act as control parameters.

By itself, the suppression of stress enhancement due to strut-breaking does not explain why at high $l/d$ and $s/d$ fracture proceeds via the breaking of struts at random positions along the line of failure.  However, it is possible that once the stress-enhancement effect has been sufficiently suppressed, so that the nearest-neighbor strut is not preferentially broken, it is  inherent material disorder which determines the strut breaking.  Any disorder due to fabrication or material inhomogeneities should be randomly distributed; this may explain why the running crack crosses over to failure by random breaking events.

\begin{figure}
\centering 
\begin{center} 
\includegraphics[width=0.85\columnwidth]{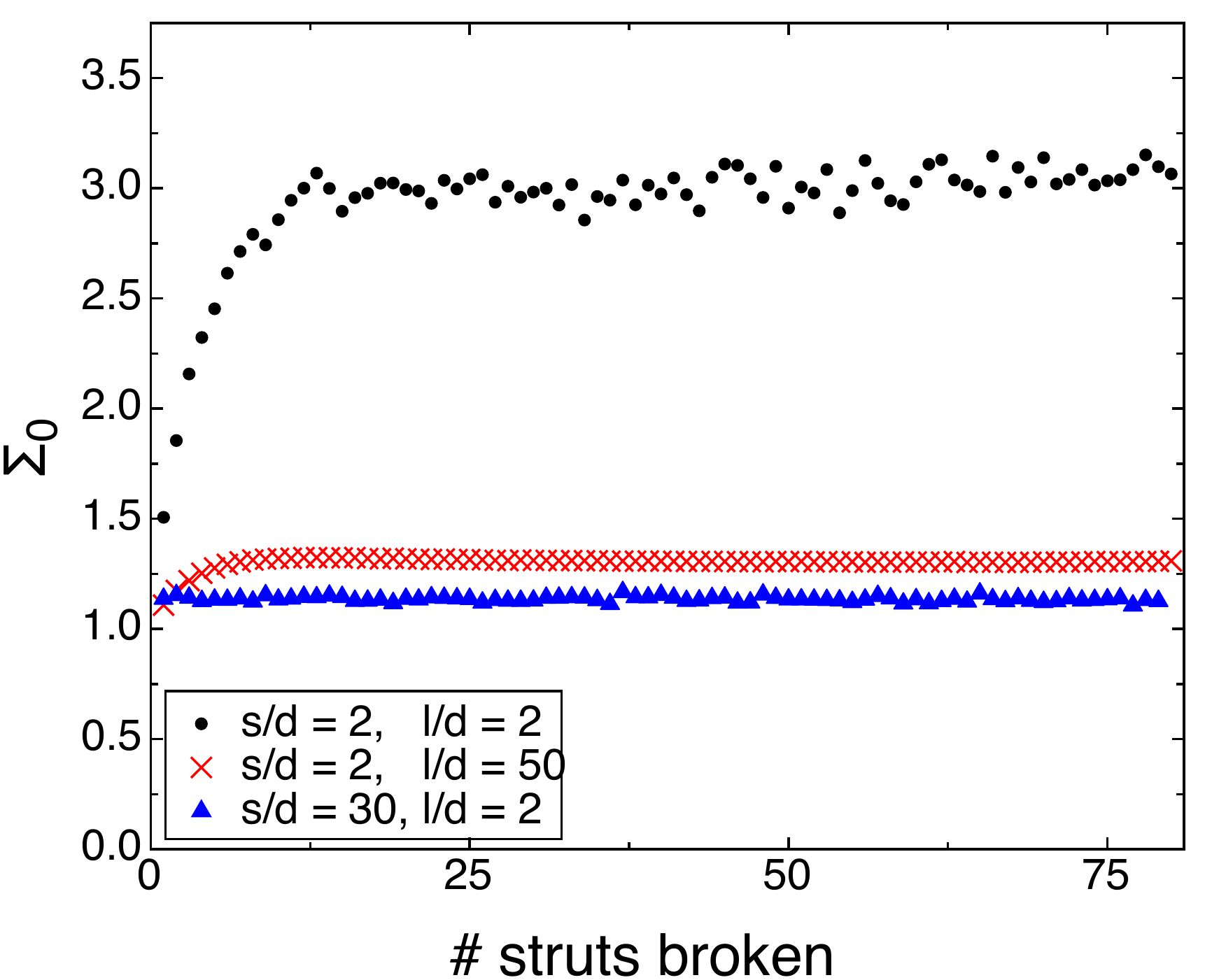} 
\caption{(Color online) $\Sigma_0$ vs.\ crack length, at low $l/d$ and $s/d$ (black circles), high $l/d$ (red X's), and high $s/d$ (blue triangles).  Only the first 80\% of breaks are shown.  In all cases after an initial increase, $\Sigma_0$ plateaus as more and more struts are broken. Furthermore, at high $s/d$ or $l/d$, $\Sigma_0$ always remains much below the value observed when $s/d$ and $l/d$ are small.} 
\label{sigma_change}
\end{center}
\end{figure}

The stress enhancement data  in Fig.\ \ref{new_stress_ex}e,f is calculated after breaking a single strut.  To investigate how $\Sigma_0$ was modified by many breaks, calculations were done which simulated an extending crack: neighboring struts were broken one-at-a-time from left to right.   Fig.\ \ref{sigma_change} shows a plot of $\Sigma_0$ versus the number of broken struts, for three samples, one in the running crack regime ($s/d = l/d =2$), and two in the random breaking regime ($s/d = 2, l/d = 50$ and $s/d = 30, l/d = 2$).  As more struts are broken, $\Sigma_0$ first increases, but then quickly plateaus.  Furthermore, at large $l/d$ or $s/d$, the stress enhancement, $\Sigma_0$, always remains much below the value at small $l/d$/$s/d$, even when many struts are broken.  Thus, the loss of stress enhancement demonstrated at high $l/d$ or $s/d$ remains throughout the breaking process.  

In conclusion, in this 1D perforated system, a transition in failure mode is observed as a function of material geometry.  The parameters $s/d$ and $l/d$ can separately be tuned to change the failure mode from dynamic (a running crack) to quasi-static (random-breaking).  Furthermore, this transition appears to be connected to the loss of stress enhancement which occurs when either $l/d$ or $s/d$ becomes large.

\section{Crack velocity as a function of geometry}
\label{velocity}

\begin{figure}
\centering 
\begin{center} 
\includegraphics[width=\columnwidth]{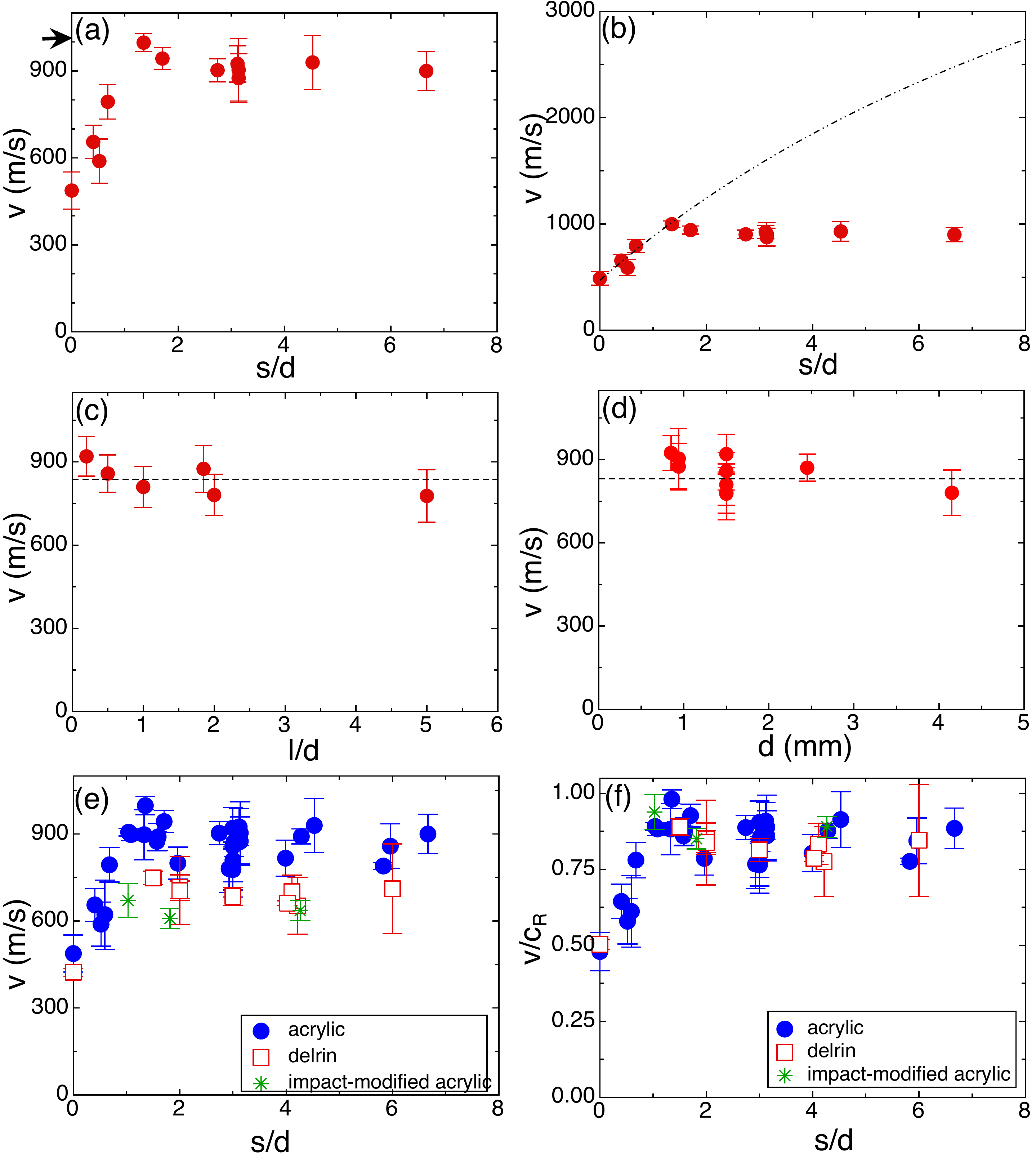} 
\caption{(Color online) (a) Crack velocity  vs.\  $s/d$.  An unperforated material is represented by $s/d = 0$.  There are two dynamical regimes:  $s/d < 1$, where the velocity increases with increasing $s/d$, and $s/d > 1$, where the velocity is independent of $s/d$.  The arrow indicates the Rayleigh wave velocity in acrylic, 1017 m/s. (b) The simple model of equation \ref{two_velocity} does not agree with the data, as indicated by the dashed line.  The fit was obtained by forcing the model to agree with the points where $s/d <$ 1.  (c)  Crack velocity vs. strut aspect ratio, $l/d$.  The crack velocity is independent of strut aspect ratio (when $l/d \lesssim 5$).  These tests were conducted for fixed $s/d$=3.  (d) Crack velocity vs.\ $d$, at a fixed value of  $s/d$=3.  The crack velocity is independent of $d$; dashed line represents the mean value of $v$.  (e) Crack velocity vs.\ $s/d$ for three different materials, acrylic (blue circles), delrin (red open squares), and impact modified acrylic (green asterisk). (f) The same data, with the crack velocity normalized by the Rayleigh wave velocity in each material, $c_R$.  The non-dimensionalized data collapse, and thus do not show any dependance on material properties. } 
\label{velocity_plots}
\end{center}
\end{figure}

As discussed in the previous section, when the $l/d \lesssim 5$ or $s/d \lesssim 10$, the 1D samples fail via a running crack; struts break in rapid succession and in an ordered manner.  Within this dynamic fracture regime, there are two velocity regimes.   At fixed $l/d$, the crack velocity, $v$, can be tuned by adjusting $s/d$, the ratio of hole size to spacing, as shown in Fig.\ \ref{velocity_plots}a.   When  $s/d < 1$, $v$ is an increasing function of $s/d$.  However, behavior changes when $s/d > 1$; $v$ saturates, and becomes independent of $s/d$.  

What sets the overall shape of the $v$ versus $s/d$ curve?  The simplest hypothesis is that the running crack moves faster because it has to create less and less free surface, e.g.\ the relative fraction of strut width to hole length is lower.  A simple model based on this hypothesis assumes that the crack travels through the struts at one velocity, $v_d$,  but then travels through the `hole region' at a different velocity, $v_s$.  Thus the crack velocity through the entire piece of perforated material, $\left<v\right>$, will be a function of both $v_d$ and $v_s$:

\begin{equation}
\left<v\right> \equiv \frac{\Delta x}{\Delta t} = \frac{d+s}{\frac{d}{v_d}+\frac{s}{v_s}} = \frac{1+s/d}{\frac{1}{v_d}+\left(\frac{s}{d}\right)\frac{1}{v_s}}
\label{two_velocity}
\end{equation}

However, this model fails completely to capture the observed behavior, as illustrated in Fig.\ \ref{velocity_plots}b.  The dashed line  was obtained by forcing equation \ref{two_velocity} to fit the points where $s/d < 1$, i.e., where the velocity appears to be increasing.  While the model can be fit to the data at low $s/d$, it predicts qualitatively incorrect behavior for $s/d > 1$: a steadily increasing velocity rather than the experimentally observed plateau.  Thus, this simple hypothesis does not explain why the crack velocity is controlled by $s/d$.

In contrast to the effect of changing $s/d$, changing the aspect ratio of the struts, $l/d$, has no effect on crack velocity.  As shown in Fig.\ \ref{velocity_plots}c, $l/d$ is varied by a factor of 25, but the crack velocity remains unchanged.  As long as the aspect ratio is small enough so that the running crack states exists ($l/d \lesssim 5$), the crack velocity is independent of $l/d$.

It should be noted that it is the \textit{ratio} of $s/d$ which controls the crack velocity, not the absolute value of $d$.  This is illustrated in Fig.\ \ref{velocity_plots}d.  The absolute value of $d$ is changed by almost  a factor of 8, yet all of the points remain within error of each other.

In this running-crack regime, it appears that crack velocities are set by geometry alone.  As a stringent test of this hypothesis, additional experiments were conducted in two other materials:  impact-modified acrylic ($Y$ = 1.76 GPa, $\rho$ = 1115 kg/m$^3$, $\nu$=0.4) and Delrin ($Y$ = 3.1 GPa, $\rho$ = 1394 kg/m$^3$, $\nu$=0.4).  As seen in Fig.\ \ref{velocity_plots}e, changing the material properties does change the absolute value of the crack velocity.  However, as shown in Fig.\ \ref{velocity_plots}f, all of the curves can be collapsed by normalizing the crack velocity by the Rayleigh velocity of each material (acrylic, $c_R$: 1017 m/s, impact-modified acrylic, $c_R$: 715 m/s, Delrin, $c_R$: 841 m/s).   This suggest that, independent of the specific details of the material, crack velocity can be controlled simply by changing the perforation geometry.

Thus, regardless of the specific failure properties of a material, a meta-material can be constructed from it which will fail in a manner determined by geometry alone.  Simply by tuning two geometric control parameters ($s/d$ and $l/d$), the failure mode can be tuned to transition from failure via a running crack to failure via the breaking of struts at random positions.  Furthermore, in the running crack regime, the crack velocity can be additionally controlled by adjusting $s/d$.

\section{Conclusion}

Failure of an elastic solid is often characterized as either ductile or brittle, i.e., either occurring via highly dissipative, slow plastic flow or by rapid, dynamic crack propagation.  Some correlations exist between material microstructure and failure mode; poly-crystalline materials have a tendency to fail ductilily \cite{garrison1987ductile}, while amorphous materials are more likely to fail in  a brittle manner \cite{griffith1921phenomena}.  However, these correlations are not one-to-one; and furthermore some materials can exhibit a transition from brittle to ductile failure as a function of temperature \cite{john1975brittle,gumbsch1998controlling}.  Clearly, the link between atomic structure and fracture behavior is nuanced.

The work presented here has explored how the geometry of a meta-material can influence how it fails under tension.  In these materials the geometry of the structure is easily accessible and tunable --- they are created simply by introducing an array of holes in an otherwise solid sheet of material.  In these perforated geometries, the transition in failure dynamics can be directly linked to changes in the underlying structure.  I have shown that the failure behavior of a perforated sheet is modified as the material becomes more sparse, and as strut bending becomes more important.  Furthermore, I demonstrate that this  transition between failure modes can be connected to a loss of stress enhancement due to the sparse geometry.  

This transition in dynamics illustrates that perforating a continuum material does more than modify its elastic constants, or change its failure strength.  Constructing a meta-material in this way allows for a new kind of material, one that has a tunable mode of failure.  Simply by choosing appropriate geometric parameters, the entire dynamics of the failure process can be adjusted.  This change in dynamics may be reminiscent of the transition seen in the fiber bundle model of 1D failure \cite{herrmann2009some, daniels1945statistical, pradhan2010failure}, where catastrophic failure failure of a complete bundle transitions to failure that occurs more locally.  However, in this model the control parameter for the transition is the amount of disorder present in the the bond breaking strengths, which is distinctly different from the system studied here.  In these experiments, geometry was varied while the amount of disorder was held \textit{fixed}.  The transition observed here is distinctly different than the one studied in these models, and thus represents a new control parameter for failure dynamics.

\section{acknowledgments} I thank Sid Nagel for his insight and support; our many fruitful discussions greatly enhanced this work.  I am also grateful to Bryan Chen, Efi Efrati, William Irvine, Noah Mitchell, Vincenzo Vitelli, and Wendy Zhang  for helpful discussions.  This work was supported by the National Science Foundation under Grant No.DMR-1404841,  the U.S. Department of Energy, Office of Basic Energy Sciences, Division of Materials Sciences and Engineering under Award DE-FG02-03ER46088, and by NSF-MRSEC under Grant No.DMR-0820054 through the use of the Chicago MRSEC Facilities. 

\bibliography{fracturebib}

\end{document}